\documentclass[twocolumn,10pt]{revtex4-1}
\pdfoutput=1

\usepackage{multibib}
\usepackage{cmbright}
\usepackage{amssymb}
\usepackage{xspace}
\usepackage{amsmath}
\usepackage{graphicx}
\usepackage[group-separator={,}]{siunitx}
\usepackage[unicode=true,pdfusetitle, bookmarks=false, breaklinks=false,pdfborder={0 0 0},backref=false,colorlinks=false] {hyperref}

\usepackage{amsfonts}

\begin{document}

\renewcommand{\figurename}{FIG.}

\title{Vortex knots in tangled quantum eigenfunctions}

\author{Alexander J Taylor}
\email{alexander.taylor@bristol.ac.uk}
\affiliation{H H Wills Physics Laboratory, University of Bristol, Tyndall Avenue, Bristol BS8 1TL, UK}

\author{Mark R Dennis}
\email{mark.dennis@bristol.ac.uk}
\affiliation{H H Wills Physics Laboratory, University of Bristol, Tyndall Avenue, Bristol BS8 1TL, UK}

\begin{abstract}\linespread{1.1} \fontfamily{cmbr}\selectfont 
  Tangles of string typically become knotted, from macroscopic twine
  down to long-chain macromolecules such as DNA.  Here we demonstrate that
  knotting also occurs in quantum wavefunctions, where the tangled
  filaments are vortices (nodal lines/phase singularities).  The
  probability that a vortex loop is knotted is found to increase with
  its length, and a wide gamut of knots from standard tabulations
  occur.  The results follow from computer simulations of random
  superpositions of degenerate eigenstates of three simple quantum
  systems: a cube with periodic boundaries, the isotropic
  3-dimensional harmonic oscillator and the 3-sphere.  In the latter
  two cases, vortex knots occur frequently, even in random
  eigenfunctions at relatively low energy, and are constrained by the
  spatial symmetries of the modes.  The results suggest that knotted
  vortex structures are generic in complex 3-dimensional wave systems,
  establishing a topological commonality between wave chaos, polymers
  and turbulent Bose-Einstein condensates.
\end{abstract}

\maketitle

Complexity in physical systems is often revealed in the subtle
structures within spatial disorder.  In particular, the complex modes
of typical 3-dimensional domains are not usually regular, and at high
energies behave according to the principles of quantum
ergodicity\cite{T,H}.  The differences between chaotic and regular
wave dynamics can be seen clearly in the Chladni patterns of a
vibrating two-dimensional plate\cite{H}, in which the zeros (nodes) of
the vibration accumulate small sand particles to appear as random
curved lines when the plate is irregular.  The spatial distribution of
these nodal lines is statistically isotropic at high energies\cite{H},
and this irregularity is typical for chaotic systems, whose ergodic
dynamics are determined only by the energy and there are no other
constants of motion.

Understanding the spatial structure of these wavefunctions can be
challenging.  Following the hydrodynamic interpretation of
single-particle quantum mechanics\cite{madelung27}, the zeros of
three-dimensional complex-valued scalar fields are, in general, lines,
which are vortex filaments around which the phase, local velocity and
probability current (in a quantum wavefunction)
circulate\cite{d,riess70,E}. This is analogous to the vortices of a
classical fluid, although the phase change around the vortex line is
quantised in units of 2$\pi$, and is singular at the vortex core where
the amplitude is zero. A similar vortex topology occurs in condensates
of many quantum particles\cite{M}.  The pattern of their vortex lines
provides a structural skeleton to wavefunctions\cite{d,E,dennis09}. In
modes above the lowest energies, the vortex pattern is far from
regular and they are densely intertwined.
A natural three-dimensional measure of the most extreme spatial
irregularity is when the vortex filaments are knotted, and here we
study the occurrence of knotted nodal vortex lines in three model
systems of wave chaos, as a natural extension of the Chladni problem
to three dimensions.

\begin{figure*}
   \centerline{\includegraphics{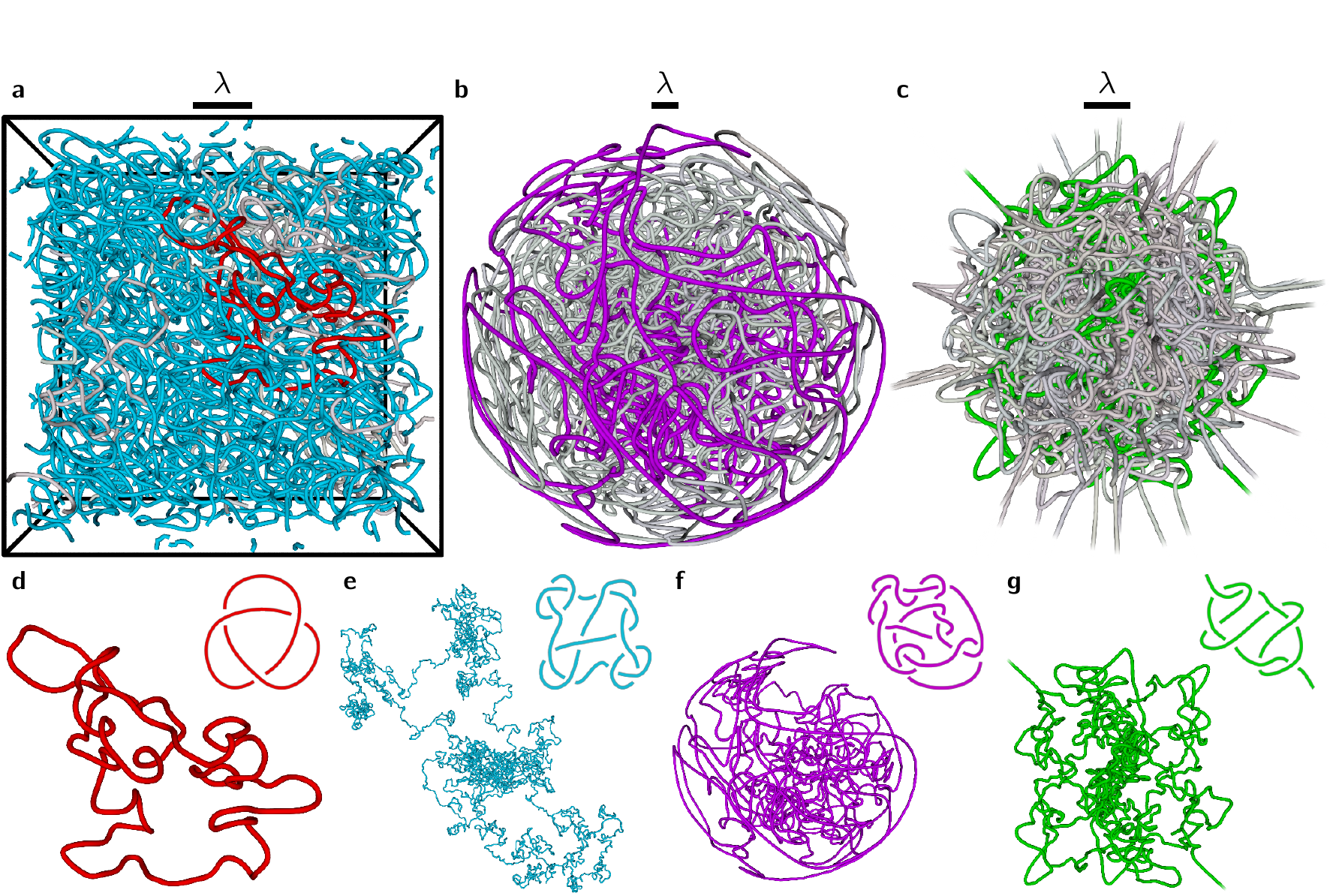}}
   \caption{ {\bf Tangled and knotted vortex filaments in random high-energy eigenfunctions of energy $E_N$. }  
   Vortex lines are shown in \textbf{a} a periodic cubic cell, with principal quantum number $N = 9$ ($E_N \propto 3N^2$); \textbf{b} the 3-sphere (plotted in a distorted projection in which all points on the bounding sphere are equivalent) with $N = 17$ ($E_N \propto N(N+2)$), and \textbf{c} the three-dimensional harmonic oscillator with $N = 21$ ($E_N \propto N+3/2$).  
   The total vortex length is similar in each eigenfunction, a reference wavelength at the origin, proportional to $E_N^{-1/2}$ is given in each of \textbf{a}, \textbf{b}, \textbf{c}.
   Each vortex loop in the eigenfunction is coloured grey except for one or two knotted examples in each system, illustrated further in \textbf{d}-\textbf{g}; each of these coloured knots is plotted alongside a simpler projection of the same knot: \textbf{d} the trefoil knot (tabulated as $3_1$ \cite{G,X}) from \textbf{a}, with length $L = 50~\lambda$ 
   and determinant $|\Delta(-1)| = 3$; 
   \textbf{e} a composite knot consisting of the two trefoils joined with the 6-crossing knot $6_2$ (i.e.~$3_1^2\#6_2$), which passes through the periodic boundary of \textbf{a} many times with $L = \num{1700}~\lambda$ and $|\Delta(-1)| = 99$;
   \textbf{f} the more complicated 14-crossing prime knot from \textbf{b}, labelled $K14n5049$ in the extended notation of standard
   tabulations beyond $11$ crossings\cite{X},
   with $L = \num{1500}~\lambda$ and $|\Delta(-1)| = 313$;  
   \textbf{g} the open 8-crossing prime knot $8_{12}$ from~\textbf{c}, having $L = \num{1000}~\lambda$ where $\lambda$ is defined with respect to momentum at the origin and only vortex length in the classically allowed region is considered, and with $|\Delta(-1)|=29$. }
   \label{fig:1}
\end{figure*}

Despite numerous investigations of the physics of diverse random
filamentary tangles, including quantum turbulence\cite{barenghi07},
loop soups\cite{Z}, cosmic strings in the early
universe\cite{vilenkin}, and optical vortices in 3-dimensional laser
speckle patterns\cite{R}, the presence of knotted structures in
generic random fields has not been previously emphasized or
systematically studied.  Vortices and defects which are knotted have
been successfully embedded in a controlled way in various
3-dimensional fields, such as vortex knots in water\cite{N}, knotted
defects in liquid crystals\cite{O,P} and knotted optical vortices in
laser beams\cite{Q}, and theoretically as vortex lines in complex
scalar fields, including superfluid flows\cite{kleckner16}, and
superpositions of energy eigenstates of the quantum hydrogen
atom\cite{bb} and other wave fields\cite{bd01}, but rigorous
mathematical techniques to resolve the statistical topology of random
fields are limited\cite{S}.  With modern high-performance computers,
the structure can be explored using large-scale simulations.

In the following, we investigate the knottedness of the nodal vortex
structures in typical chaotic eigenfunctions of comparable size
(i.e.~energy and total nodal line length) for three model systems.
The chaotic eigenfunctions are represented as superpositions of
degenerate energy eigenfunctions weighted by complex, Gaussian random
amplitudes; such superpositions are established as good models of wave
chaos in the semiclassical limit of high energy \cite{F,H,T}, since
the wave pattern is determined only by their energy (unlike a plane
wave which has a well-defined momentum).  Our three systems are the
cubic cell with periodic boundary conditions, whose degenerate
eigenfunctions are plane waves; the abstract 3-dimensional sphere
(3-sphere), whose degenerate eigenfunctions are hyperspherical
harmonics and which has finite volume but no boundary; and the
isotropic, 3-dimensional harmonic oscillator (3DHO), whose nodal
structure is largely contained within the classically allowed region
where energy exceeds the potential, such as could describe an
isotropic harmonically-trapped Bose-Einstein
condensate\cite{L}. Further information and illustrations of these
eigenfunction systems appear in Supplementary Note 1 and Supplementary
Figures 1-4.

We find that vortex knots occur with high probability at sufficiently
high energy in all of these random wave superpositions.  Even in low
energy cavity modes, possibly accessible to experiments, the results
suggest knotted vortices will occur with some reasonable probability.
The statistical details of random vortex knotting (with respect to the
types of knot that occur, the length of knotted vortex curves, and the
eigenfunction energy) depends strongly on the wave system, and we
perform an analysis and comparison of these properties.

\section*{Results}

\subsection*{Knots in high energy random eigenfunctions}

Figure~\ref{fig:1} shows the nodal/vortex structure of a typical
chaotic eigenfunction in each model system: in 1a, a cube with
periodic boundaries; in 1b, the 3-sphere; and in 1c, the 3DHO.  In all
three cases the random modes are labelled by energy $E_N$ with
principal quantum number $N$ (further details of how the model systems
are generated and how the vortices are located is given in
Supplementary Notes 1 and 2).  In each of Figures 1d-g, a single vortex line
from these eigenfunctions is shown alongside a simpler projection of
the same knot.  Each vortex line resembles a random walk \cite{U,R}.

Our analysis of these knots uses the standard conventions of
mathematical knot theory\cite{G}, in which knots in closed curves can
be factorised into prime knots which are tabulated according to their
minimum number of crossings (examples are shown in Supplementary
Figure 5).  Both prime knots (e.g.~as Figure~1d,f-g), and composites
of prime knots (e.g.~as Figure~1e), are identified in the curve data
by a combination of several schemes.  After simplifying the longest
curves by a geometric relaxation method\cite{U}, several topological
invariants are computed for each curve, which distinguish knots of
different types.  These are: the absolute value of the Alexander
polynomial\cite{k} $|\Delta(t)|$, evaluated for $t$ at certain roots
of unity; the hyperbolic volume\cite{V}; and the second and third
order Vassiliev invariants\cite{J,W}.  Although other knot invariants
can have more discriminatory power than these separately~\cite{G,X},
they tend to be significantly more demanding in computer time: in
practice the combination of this particular set of invariants
discriminates almost all tabulated knots\cite{G,X} (up to at least 14
crossings; the methods are described further in Supplementary Note 3).
Each of these invariants encodes different topological and geometric
information about each knot\cite{G}.  When comparing the knottedness
of different vortex loops, we follow previous studies\cite{k,J} in
using the (positive integer-valued) knot determinant $|\Delta(-1)|$ as
the primary quantitative measure of their knotting complexity.

We have identified approximately \num{50000} knots from around $10^9$
curves (of different lengths), computed from about $10^6$ random
eigenfunctions (at different energies).  Some of these knotted curves
are relatively short (such as that in Figure~1d), although the
majority of vortex knots are much longer, such as in Figure~1e, whose
full spatial extent spreads over several periodic cells.  At large
lengthscales, each curve approximates a Brownian random
walk\cite{U,R}; for closed curves, the radius of gyration scales with
the square root of the total length.  Most of the vortex knots occur
in closed loops; in the 3DHO, however, most vortex lines stretch to
spatial infinity as almost-straight lines in the classically forbidden
region.  Most of the knots in the 3DHO are in these open curves, such
as that represented in Figure~1g.  In the periodic cube, many of the
longer lines wrap around the cell a nonzero number of times (i.e.~they
have nontrivial homology, so in a tiling of the 3D cells they would be
infinitely long and periodic\cite{R}).  Since knots in such a periodic
space have not been adequately classified mathematically\cite{G}, we
only consider knots in closed loops with trivial homology in this
system.

Links, which are configurations of two or more vortex curves that are
topologically entangled with one another, also occur frequently.  In
fact we find links to be more common than knots, consistent with
previous investigations of random optical fields at smaller
lengthscales in which links are found to be common but knots were not
detected\cite{ae}.  The restriction to knotting in the present study
was chosen as it allows comparison with extensive studies of the
knotting probability of random curves (as random walks)\cite{k},
whereas the study of random linking is not so well developed.
Furthermore, linking of open curves (in the 3DHO) is not well-defined.
For these reasons, our analysis is limited to the statistics of vortex
knotting.

\subsection*{Probabilities and complexities of knottedness}

\begin{figure}
   \centerline{\includegraphics{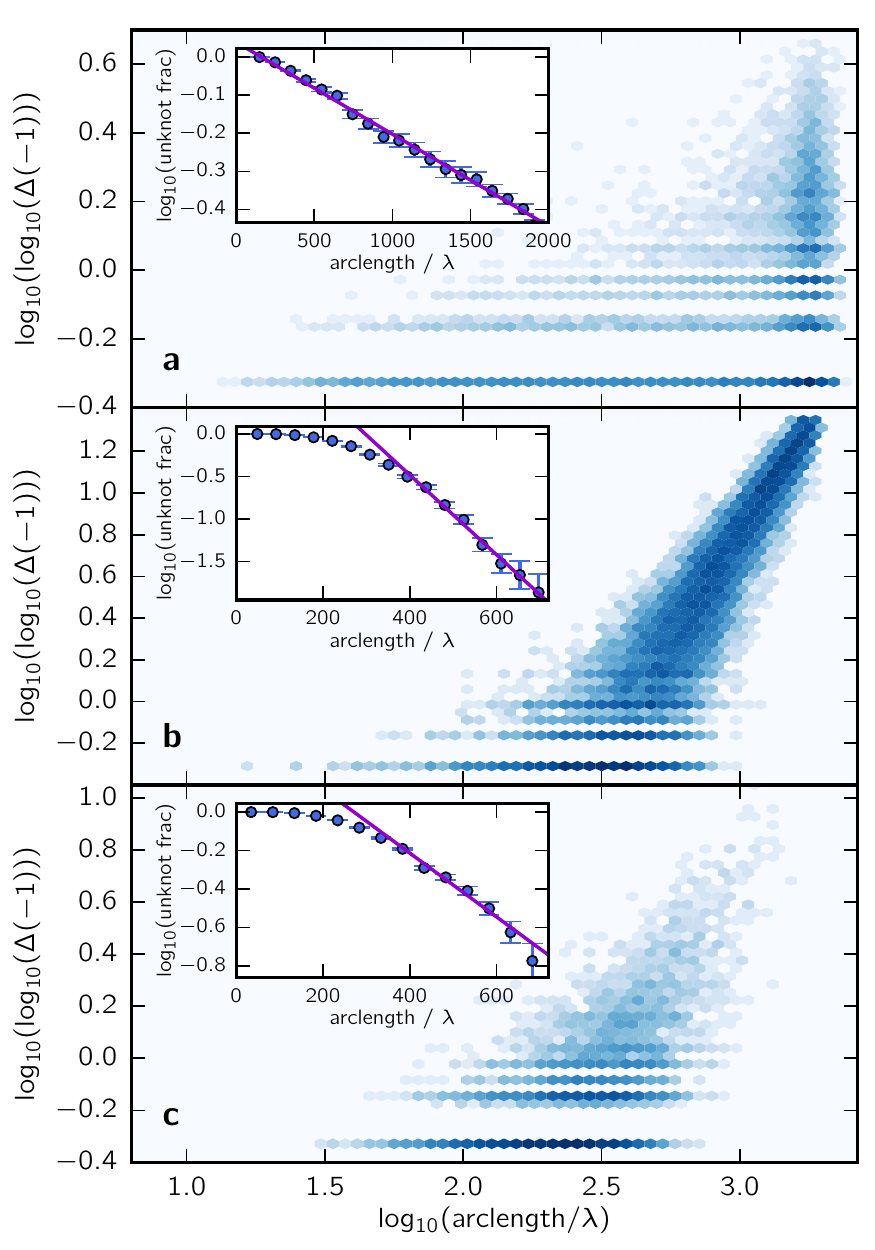}}
   \caption{
     \textbf{Dependence of vortex knot complexity on curve length.}
     The plots show histograms of $\log_{10}(\log_{10} |\Delta(-1)|)$ against $\log_{10}(L)$ for the thee systems at the same energies of the examples in Figure 1: 
     \textbf{a} periodic cube, $N=9$; \textbf{b} 3-sphere, $N=17$; \textbf{c} three-dimensional harmonic oscillator, $N=21$.
     In each system, the logarithm of the probability that a given vortex curve is unknotted as a function of its length $L$ is plotted in the insets, with error bars representing the standard error on the mean probability over many different random eigenfunctions. 
     For larger values of $L$, this unknotting probability logarithm is fitted to $-L/L_0 + \mathrm{const}$, with scaling factors $L_0$ of approximately: \textbf{a} $\num{1800}~\lambda$; \textbf{b} $100~\lambda$; \textbf{c} $250~\lambda$.  
     In the main plots, the strong horizontal lines of constant, low-value $|\Delta(-1)|$ correspond to specific knots, such as the trefoil (for which
     $\log_{10}(\log_{10} |\Delta(-1)|) = -0.32$) and the composite double trefoil $3_1\#3_1$ (for which $\log_{10}(\log_{10} |\Delta(-1)|) = -0.02$).
   }
   \label{fig:2}
\end{figure}

\begin{figure}
   \centerline{\includegraphics{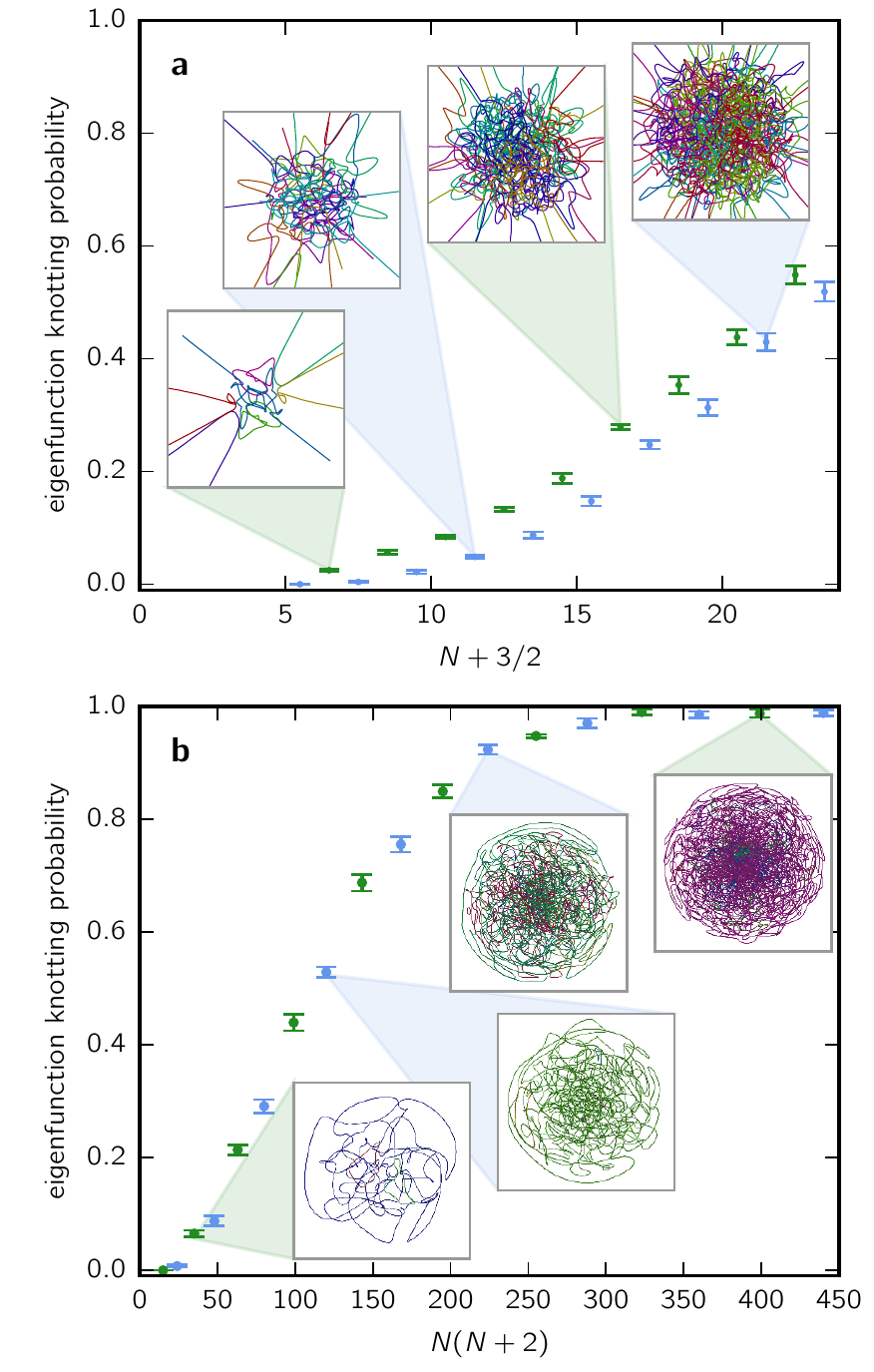}}
   \caption{ 
   {\bf Probabilities that, as functions of energy, a given random eigenfunction in the system contains knotted vortices.}  
   \textbf{a} three-dimensional harmonic oscillator; \textbf{b} 3-sphere. 
   Blue (green) points denote those where the principal quantum number $N$ is even (odd), for reasons described in the main text, and errors represent the standard error on the mean fraction of knotted eigenfunctions, averaged over many different random samples. 
   Insets depict the vortices in a typical eigenfunction at different energies, with each vortex curve is represented in a different colour.  }
   \label{fig:3}
\end{figure}

\begin{figure*}
   \centerline{\includegraphics{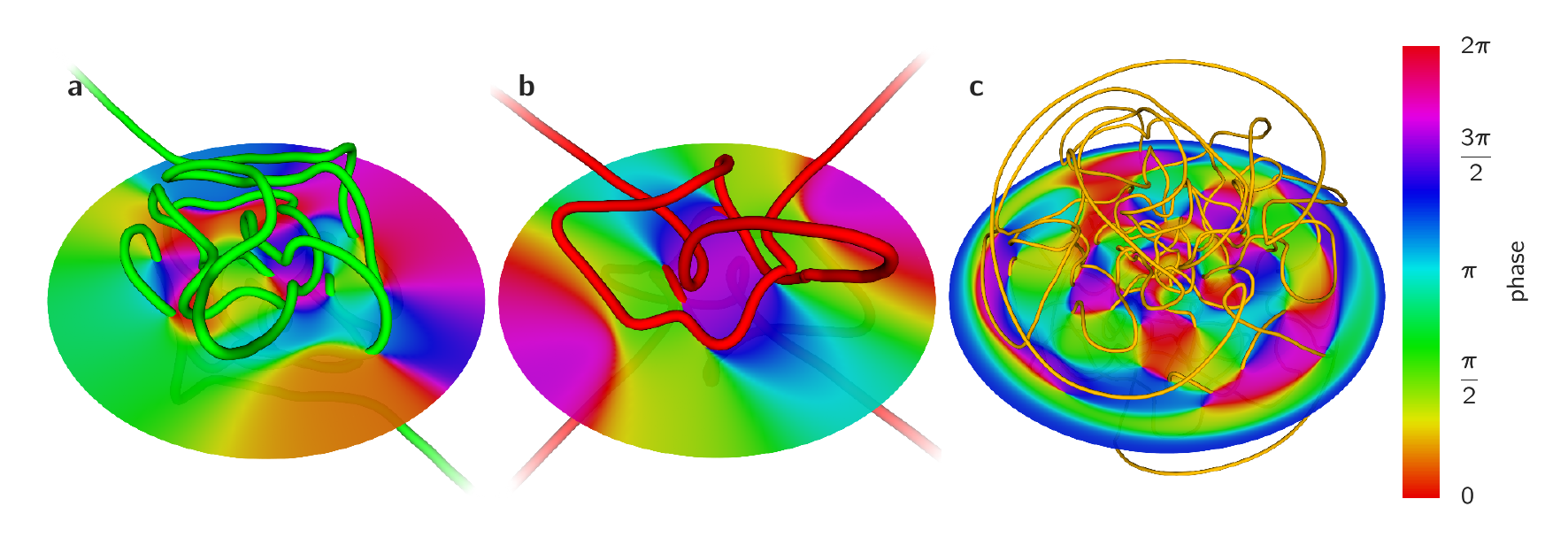}}
   \caption{ {\bf Examples of low-energy eigenfunctions with knotted vortices in symmetric strongly amphicheirally symmetric conformations.}
     \textbf{a} three-dimensional harmonic oscillator (3DHO), $N = 5$, with a single open figure-8 vortex knot (tabulated as $4_1$) in a strongly negatively amphicheiral conformation (this is the simplest prime knot with this symmetry);
     \textbf{b} 3DHO, $N = 4$, with a pair of mirrored trefoil vortex knots $3_1\#3_1$ (knotting at this low energy is found to be extremely rare);
     \textbf{c} 3-sphere, $N = 7$, with the knot $10_{99}$ (the simplest prime knot with strongly positive amphicheiral symmetry\cite{AA}), but omitting the two other small vortex loops occurring in this eigenfunction.  
     In each case, the phase in a plane through the centre of the complex random wavefunction is represented, coloured by hue. 
     Supplementary Movies 1-3 show these wavefunctions from varying viewpoints.}
   \label{fig:4}
\end{figure*}

As one might expect, the random eigenfunction statistics show that longer knotted vortex curves display a greater complexity of knot types, as measured by the knot determinant $|\Delta(-1)|$.  
This is represented in Figure~2 for each of the three systems at fixed energy.
These histograms indicate that the distribution of $\log|\Delta(-1)|$ with respect to vortex length $L$ apparently scales according to a power law.  
Longer curves are also more likely to be knotted; the probability of a given vortex loop being unknotted decreases exponentially according to $L$ (Figure~2 insets), as previously studied for random walks modelling polymers\cite{I,J,k}.  
The value of the unknotting exponents is different for each of the three systems, given in the caption.

Each of the model systems studied has a finite spatial extent: the side length of the cube, and the diameters of the 3-sphere and classically-allowed region of the 3DHO.  In the latter two cases, this finite size imposes a long-length cutoff $S$ to the Brownian scaling of long vortex lines.  
In the 3-sphere and 3DHO, loops whose radius of gyration exceeds $S$ are confined, and the Brownian and confined regimes have different knotting probability exponents.  
Both knotting probability and complexity increase distinctly with length in the confined regime, where long curves are restricted to volumes much smaller than the corresponding Brownian radius of gyration would allow.

In contrast, the cube's periodic boundary conditions allow the vortex lines to have extent greater than the cube side length, although the Brownian scaling of loop sizes gives way at some $S$ to the lines with nontrivial homology, which do not contribute to the knot count.
Figure~2b,c shows that the minimal complexity of knots in the confined regime also scales with $L$; long vortices are always knotted in these systems, and the knot is usually very complex.  
By contrast, under periodic boundary conditions (Figure~2a) even the longest vortex loops can be topologically trivial.  
In all of these cases, the scaling depends weakly on the energy of the eigenfunction, at least in the ranges of energy considered in the simulations.

\subsection*{Low energy knots and the impact of eigenfunction symmetry}

The total vortex length in random eigenfunctions is proportional to their energy $E_N$ (see Supplementary Note 1).  
It is therefore natural to expect knotted vortex lines to occur more frequently in higher-energy eigenfunctions.  
This probability of knotting with energy is shown in Figure~3 for eigenfunctions of the 3-sphere and 3DHO.  
In the 3-sphere, where all vortex loops must be closed, this probability reaches 99\% when $N \approx 16$, which is surprisingly small to guarantee such a degree of topological complexity (as illustrated in the insets).

The energies simulated in the 3DHO are insufficient to guarantee that
a random eigenfunction will contain a knotted vortex (Figure~3a), as
the distribution of vortex lengths in the classically allowed region
prefers a greater number of shorter curves.  The knotting probability
of a 3DHO eigenfunction also strongly depends on whether the principal
quantum number $N$ is odd or even, since 3DHO energy eigenstates, and
hence the zeroes (i.e.~the vortex tangle), must be parity-symmetric
with respect to spatial inversion through the origin; in fact, the
nodal system of vortex curves is required to be strongly negatively
amphicheiral\cite{AA} (i.e.~the tangents at parity-opposite vortex
points must be parallel).  Furthermore, when $N$ is odd, exactly one
open vortex line must pass through the origin, such as the low-energy
example in Figure~4a.  Since strongly negatively amphicheiral knots
must be open (see Supplementary Note 4 and Supplementary Figure 6),
this line can take such a configuration and is commonly knotted,
contributing to a larger knotting probability when $N$ is odd (other
knots can occur when $N$ is odd or even, but only in antipodal pairs,
as in Figure~4b).  This symmetry constraint applies to all energies
considered in the simulations.

Eigenfunctions of the 3-sphere must also be symmetric under inversion
between antipodal points, but here the curve system must be strongly
positively amphicheiral (opposite vortex points have antiparallel
tangents).
The simplest prime knot with this symmetry is in fact the 10-crossing knot
$10_{99}$\cite{AA}, and an example of a knotted vortex of this type is shown in Figure~4c.

\section*{Discussion}

Based on computer experiments, we have shown that knotted vortex lines
are common in random quantum wavefunctions, even at comparatively low
energies.  These results complement previous rigorous mathematical
studies where small knots (of any type) have been proven to occur with
nonzero probability in random eigenfunctions of the 3DHO\cite{ac} and
hydrogen atom\cite{peraltasalashydrogen} at high energies.  These
knots would be quite different from those studied in the present work,
as they occupy limited volumes and, in the 3DHO, occur with
exponentially small probabilities.  As such, those knots may be
thought of as highly structured superoscillatory
phenomena\cite{berry94}, known to be very rare.  On the other hand,
the knotted vortices emphasised here can extend throughout the entire
wavefunction, and are rather common except at the lowest energies; in
fact, the very smallest knots we find in the 3DHO are an order of
magnitude larger than the bounds on the size of superoscillatory examples.
Preliminary numerical studies of vortex knots in random
eigenfunctions of hydrogen indicate similar behaviour to the 3DHO
described here, where eigenfunctions containing knotted vortices are
typical beyond a certain energy (and comparable mode count), although
unlike the 3DHO and 3-sphere, the nodal sets are not constrained by an
amphicheiral symmetry.

Although our computer experiments have been limited to modes of fixed
energy, we do not anticipate major differences in systems of random
waves with different power spectra, such as a turbulent Bose-Einstein
condensate\cite{L} with characteristic power law spectrum, especially
when sizes of system are similar to or greater than those considered
here.

Knotting of vortices in chaotic complex 3D eigenfunctions is a complex
counterpart to Chladni's original observation of the structure of
cavity modes, and we anticipate that generic 3D complex cavity
modes\cite{H,T} will contain knotted vortices.  Our results lead us to
expect that almost every random quantum eigenfunction at sufficiently
high energy contains at least one knotted vortex line and, at least in
the systems considered here, this is common even at the relatively low
energies currently accessible to numerical experiment.  Knotted nodal
lines with characteristic complexity scalings may therefore be a
complex, 3-dimensional counterpart to the nodal statistics proposed to
signify quantum chaos\cite{ad} for real-valued chaotic eigenfunctions
in two dimensions.  Based on the mode counts\cite{H} of the energies
in Figure~3, we expect knotted vortices to occur with 50\% probability
from somewhere between the 500th (3-sphere) to \num{3000}th (3DHO)
mode of a chaotic cavity.

\subsection*{Methods}

\textbf{Sampling random eigenfunctions.}  We generate random
eigenfunctions of model systems via standard random superpositions of
their degenerate eigenfunctions.  Further details and comparisons of
these systems, especially their eigenfunctions, are included in 
Supplementary Note 2.

\textbf{Tracking vortices (nodal lines).}  Within each random
eigenfunction, phase vortices are numerically tracked by sampling the
wavefunction on a cubic lattice (initial voxel size approximately
$0.1\lambda$) and checking around each numerical grid plaquette for
the circulation of phase indicating the passage of a vortex line
through the face of the grid cell.  The phase is only guaranteed to
circulate in this way when the amplitude is zero but the complex
scalar gradient is non-zero (i.e.~the vortices occur as transverse
intersections of real and imaginary nodal surfaces), but this is
generically the case in random eigenfunctions; the algorithm would
otherwise fail to converge on the vortex line but, as anticipated, we
have never observed this to happen. This core procedure does not
always fully capture the geometry of the vortex curve (especially
where two vortex lines approach closely), and so the above procedure
is augmented with a recursive resampling of the numerical grid, at
successively higher resolutions, in any location where the vortex line
tracking is ambiguous; we call this the RRCG algorithm (see
Supplementary Note 2).  Since vortex lines must be continuous,
the result of the algorithm is a piecewise-linear representation of
each vortex in the entire large scale tangle.  The vortex topology is
retained, and geometry well recovered, even with a relatively poor
initial sampling resolution.  The algorithm generalises to each of
periodic boundary conditions, the 3-sphere and the 3DHO via an
appropriate choice of lattice boundary conditions, and in the case of
the 3-sphere by sampling on multiple Cartesian lattices and joining
appropriate boundaries to reproduce the geometry of the sphere.

\textbf{Detecting and classifying knots and links.}  The topology of
vortices is investigated using the standard knot invariants described
in the main text, whose values depend only on the topology of a single
curve and from which the knot type can be inferred.  It is
computationally expensive to use some invariants known to be more
powerful (e.g.~the Jones or HOMFLY-PT polynomials), or to identify the
exact knot type of every curve.  When necessary (Figure 1, Figure 4)
we are able to do so unambiguously using the Alexander polynomial
(evaluated at certain roots of unity), hyperbolic volume, and
Vassiliev invariants or degree 2 and 3.  We further use the knot
determinant to measure the topological complexity of vortex curves
even without identifying their specific knot types.

\subsection*{Data availability}

All relevant data are available from the corresponding author upon request.

\subsection*{Acknowledgments}
The authors thank Michael Berry for valuable discussion and for
originally proposing the problem investigated here, and Andrea Aiello,
Stu Whittington, Daniel Peralta-Salas and Dmitry Jakobson for useful
suggestions. This research was funded by a Leverhulme Trust Research
Programme Grant. This work was carried out using the computational
facilities of the Advanced Computing Research Centre, University of
Bristol. The authors are grateful to the KITP for hospitality during
the early part of this work.  Alexander Taylor was partially funded by
the Engineering and Physical Sciences Research Council and Mark Dennis
by the the Royal Society of London.
  
\subsection*{Author contributions}

A J Taylor created the numerical routines for simulating
eigenfunction modes and performing topological analysis. Other
contributions were shared equally.

\clearpage

\appendix

\renewcommand{\figurename}{SUPP. FIG.}
\setcounter{figure}{0}

\section*{Supplementary Note 1:\\ Random wave formulation and nodal statistics}

The wavefunctions whose nodal structures are considered in the main
text are random superpositions of degenerate energy eigenstates in a given system,
considered over 3-dimensional position $\vec{r} = (x,y,z)$,
\begin{equation}
   \psi_j^N(\vec{r}) = \sum_j a_j \Psi^N_j(\vec{r})
   \label{eq:psidef}
\end{equation}
where the sum is over a finite set of indices labelled by $j$, the
$a_j$ are Gaussian random complex variables, and the $\Psi^N_j$
satisfy the time-independent Schr\"odinger equation
$\hat{H} \Psi^N_j = E \Psi^N_j$ for some 3-dimensional Hamiltonian
operator $\hat{H}$.  Thus $\hat{H} \psi_N = E \psi_N$, and $N$
denotes an integer quantum number.

In the usual random wave model (RWM) which is taken to model wave
chaos (for instance, quantum chaotic eigenfunctions in the
semiclassical limit)~\cite{F,L}, 
the Hamiltonian is
$\hat{H} = -\tfrac{\hbar^2}{2M}\nabla^2$ for a single particle of
mass $M$ at a high energy, so the sum over $j$ is effectively
infinite. The ensemble of random functions is statistically
isotropic, homogeneous and ergodic, and the (non-normalizable) basis
states $\Psi_j$ can be taken to be plane waves with the same spatial
frequency $\Psi_j(\vec{r}) = \exp(i \vec{k}_j\cdot \vec{r})$, where
$E = \tfrac{\hbar^2}{2M}|\vec{k}_j|^2$.

Unlike the RWM, our numerical realisations are systems involving
superpositions over a finite number of degenerate energy
eigenfunctions (indexed by principal quantum number $N$), whose
spatial complexity only occupies a finite spatial volume yet whose
spatial configuration (including the vortex lines) is statistically
similar to the RWM.  These systems are the periodic cubic cell, the
3-sphere and the isotropic three-dimensional harmonic oscillator
(3DHO). The statistical behaviour of the eigenfunctions of each of
these systems approaches that of the isotropic RWM in the limit of
high energy.  We compare these systems at a range of different
energies, from those where knotted vortices first appear to the
highest energies practically accessible using the RRCG algorithm introduced in 
Section \ref{sec:numericaltechniques}.  Sample functions of the
two-dimensional analogues of these complex random fields are shown in
Supplementary Figure \ref{fig:2dsamples}, where the vortices occur as points (nodes
of modulus, phase singularities).

The periodic 3-cell (flat 3-torus) of side length $L$ has the most
direct connection to the treatment of the infinite bulk RWM.  The
Hamiltonian is again $-\tfrac{\hbar^2}{2M}\nabla^2$ and the
eigenfunctions are plane waves with Cartesian components
proportional to integers, $\vec{k}_j = \tfrac{2\pi}{L}(\ell,m,n)$.
The corresponding energy is
$\tfrac{2\pi^2\hbar^2}{ML^2}(\ell^2 + m^2 + n^2)$, and the
degeneracies follow naturally from different triplets of integers
having the same sum of squares, with $j$ acting as an index
over such triplets.

An extra consideration determines which eigenfunctions of the periodic cell
are chosen for our study of vortex tangling.  For a typical
eigenfunction, the complex
field $\psi_N$ is periodic with a cubic fundamental cell.  However, it
is not difficult to show that for any such $\psi$, if
$\psi(\vec{r}) = 0$, then
$\psi\left(\vec{r}+\tfrac{L}{2}(1,1,1)\right) = 0$ and hence the
periodicity of a typical eigenfunction's nodal structure is
body-centred cubic. The primitive cell of such a lattice, and
therefore of the nodal line tangle, is a truncated octahedron. For
simplicity in numerically tracking vortices through the periodic
boundaries, we prefer to describe a periodic nodal structure whose
primitive cell is a cube. These symmetries of nodal lines in a larger
field with cubic symmetry are illustrated in Supplementary Figure
\ref{fig:periodic}.

Certain energies give rise to extra symmetries which guarantee this
property.  When energies are chosen to be
$E = \tfrac{2\pi^2\hbar^2}{ML^2} 3 N^2$ for integer $N$, that is,
$\ell^2 + m^2 + n^2 = 3N^2$, then $\ell, m, n$ must be all odd or all
even, depending on whether $N$ is odd or even.  The nodal structure of
a superposition of these plane waves has a primitive cell which is
cubic with side length $L/2$, which is an octant of the original cubic
cell of the complex wavefunction.  It is these smaller cells which are
considered in the main text.  Energies $E_N \propto 3N^2$ are
guaranteed to have at least an eightfold degeneracy with
$(\ell,m,n) = (\pm N, \pm N, \pm N)$, and in practice (for
sufficiently high $N$) the degeneracy is much higher. In the examples
in the main text, $N$ is chosen to be $9$, so triplets of integers
whose sum of squares equals 243 are 3,3,15; 5,7,13; 1,11,11 as well as
9,9,9.  All together, the total number of plane waves at this energy
(counting permutations and all possible signs of components) is 104.
The nodal statistics for random eigenfunctions at this energy closely
recover local geometrical statistics expected of the isotropic
model~\cite{U}.
 
In the 3-sphere, coordinates are specified in terms of three angles,
$\chi,\theta,\phi,$ with
$0 \le \chi, \theta \le \pi, 0 \le \phi < 2\pi$.  The energy
eigenfunctions are those of the (normalised) Laplace-Beltrami operator on the
3-sphere, which are the \emph{hyperspherical harmonics}~\cite{hochstadt71},
\begin{widetext}
  \begin{equation}
    \mathcal{Y}_{N\ell m}(\chi, \theta, \phi) = \sqrt{\frac{2^{2\ell+1} (N-\ell)!
        (1+N)}{\pi(1+\ell+N)!}} \,\ell !\, \sin^\ell(\chi)\, C_{N-\ell}^{(\ell+1)}(\cos \chi)\,
    Y_\ell^m(\theta, \phi)~,
  \end{equation}
\end{widetext}
where $Y_\ell^m$ are the usual spherical harmonics of the 2-sphere,
$C_{N-\ell}^{(\ell+1)}$ are \emph{Gegenbauer polynomials}, and for
integers $N,\ell,m$, $0 \le N$, $0 \le \ell \le N$ and
$-\ell \le m \le \ell$.  The corresponding eigenvalues are labelled by
the principal quantum number $N$, with $E_N = N (N+2)$ (in appropriate
units) which are therefore $(N+1)^2$-fold degenerate with the label
$j$ corresponding to different values of $\ell$ and $m$.  In the main
text, nodal structures are calculated in these systems for $N$ up to
21. In the stereographic projection of the 3-sphere into spherical
polar coordinates, $\theta$ and $\phi$ are the usual spherical angles,
and the radial coordinate is $\tan(\chi/2)$. 

Random waves in the 3DHO are randomly weighted degenerate
eigenfunctions of the Laplacian with an isotropic harmonic potential
with angular frequency $\omega$,
$$\hat{H}\psi_N = -\tfrac{\hbar^2}{2m}\nabla^2 \psi_N(\vec{r}) + \tfrac{m \omega^2}{2}\vec{r}^2\psi_N(\vec{r}) = E_N\psi_N(\vec{r})~,$$
where $E_N = \hbar \omega (N + \tfrac{3}{2})$ for integer $N \ge 0.$
Following the standard theory of the three dimension isotropic
harmonic oscillator, the energy eigenfunctions are
$\tfrac{1}{2}(N+1)(N+2)$-fold degenerate~\cite{rae07}.
Multiple different bases of energy eigenfunctions can be chosen, but
the simplest option for numerical computation utilises Hermite
polynomials $H_n$ arising from separation of variables in Cartesian
coordinates,
\begin{widetext}
  \begin{equation}
    \Psi^N_j(\vec{r}) = \frac{1}{\sqrt{2^N \ell!m!n!}} \left(\frac{M \omega}{\pi \hbar}\right)^{3/2} 
    H_{\ell}\left(\sqrt{\tfrac{M \omega}{\hbar}} x\right) 
    H_m\left(\sqrt{\tfrac{M \omega}{\hbar}} y\right)
    H_n\left(\sqrt{\tfrac{M \omega}{\hbar}} z\right)
    \exp\left(-\frac{M \omega}{2\hbar}[x^2+y^2+z^2]\right)~,
  \end{equation}
\end{widetext}
where $j$ labels triples of nonnegative integers $\ell,m,n$ such that $N = \ell + m + n$.
The resulting vortex tangle is largely confined to within the
classical radius $r=\sqrt{2E}$, outside which vortices quickly become
geometrically trivial, although they may extend infinitely as shown in
Supplementary Figure \ref{fig:qho_classical_region}.

\begin{figure}
  \centering{
  \includegraphics{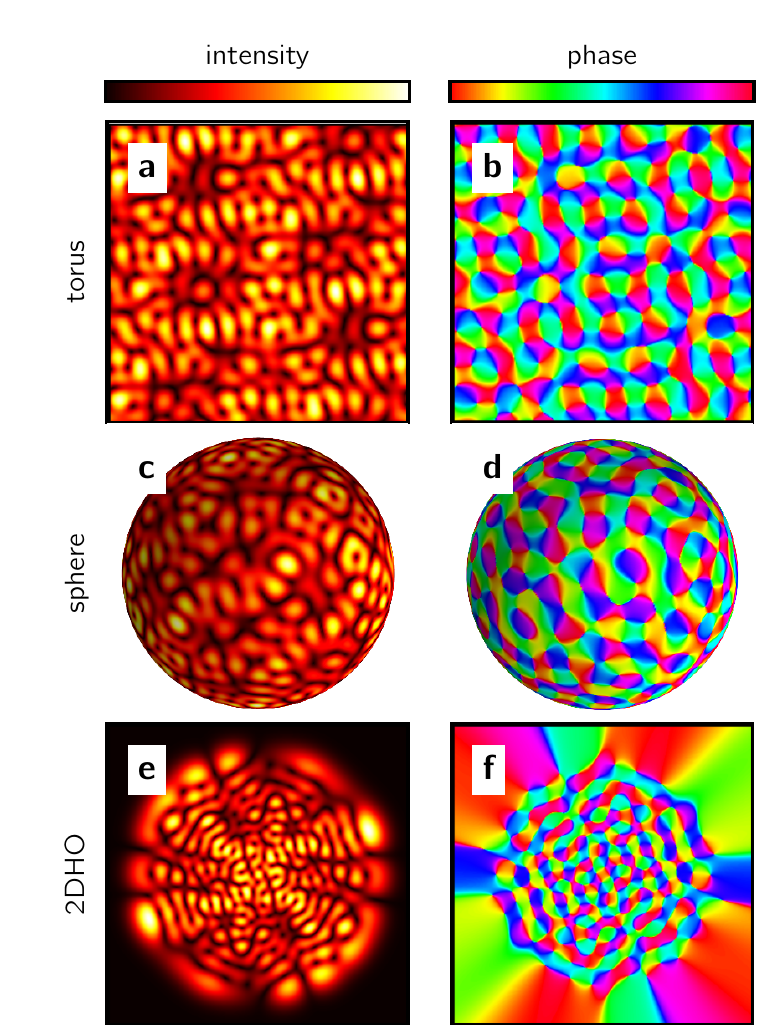}
  }
  \caption{The intensity (modulus squared) and phase (argument) of sample random energy eigenfunctions in 2D systems. 
  \textbf{a}-\textbf{b} square with periodic boundary conditions (flat 2-torus); \textbf{c}-\textbf{d} the 2-sphere; \textbf{e}-\textbf{f} the 2D harmonic oscillator (2DHO).
  These are found using 2D analogues of (\ref{eq:psidef}), i.e.~complex random superpositions with the same energy (spatial frequency) of 2D plane waves in \textbf{a} and \textbf{b}, spherical harmonics in \textbf{c} and \textbf{d}, and 2D harmonic oscillator eigenstates (Hermite-Gauss functions) in \textbf{e} and \textbf{f}.
}
  \label{fig:2dsamples}
\end{figure}

\begin{figure*}
  \centering{
  \includegraphics{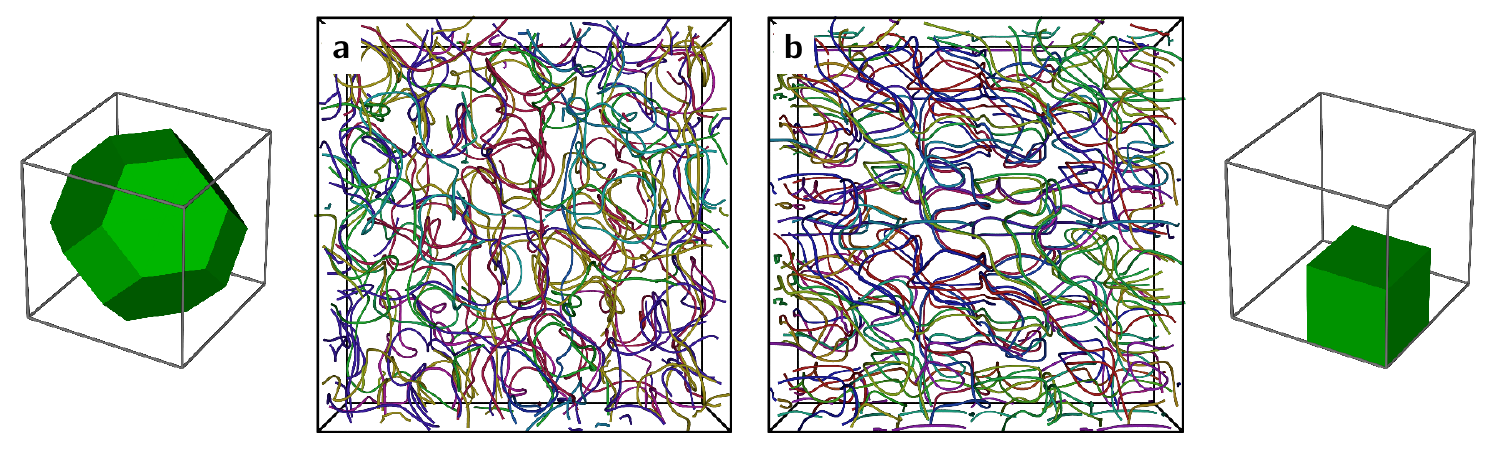}
  }
  \caption{ Vortices in random energy eigenfunctions in a periodic
    cubic cell; \textbf{a} without and \textbf{b} with the additional constraint
    to octant symmetry.  In \textbf{a}, $\ell^2+m^2+n^2=26$ and the periodic
    unit cell of the vortex tangle is the truncated octahedron. In
    \textbf{b}, $\ell^2+m^2+n^2=27 = 3\times 3^2$ and the additional symmetry
    due to all $\vec{k}$ components being odd means that the periodic
    unit cell of the vortex tangle is an octant of field's primitive
    cubic cell.}
  \label{fig:periodic}
\end{figure*}

\begin{figure}
  \centering{
  \includegraphics[width=0.4\textwidth]{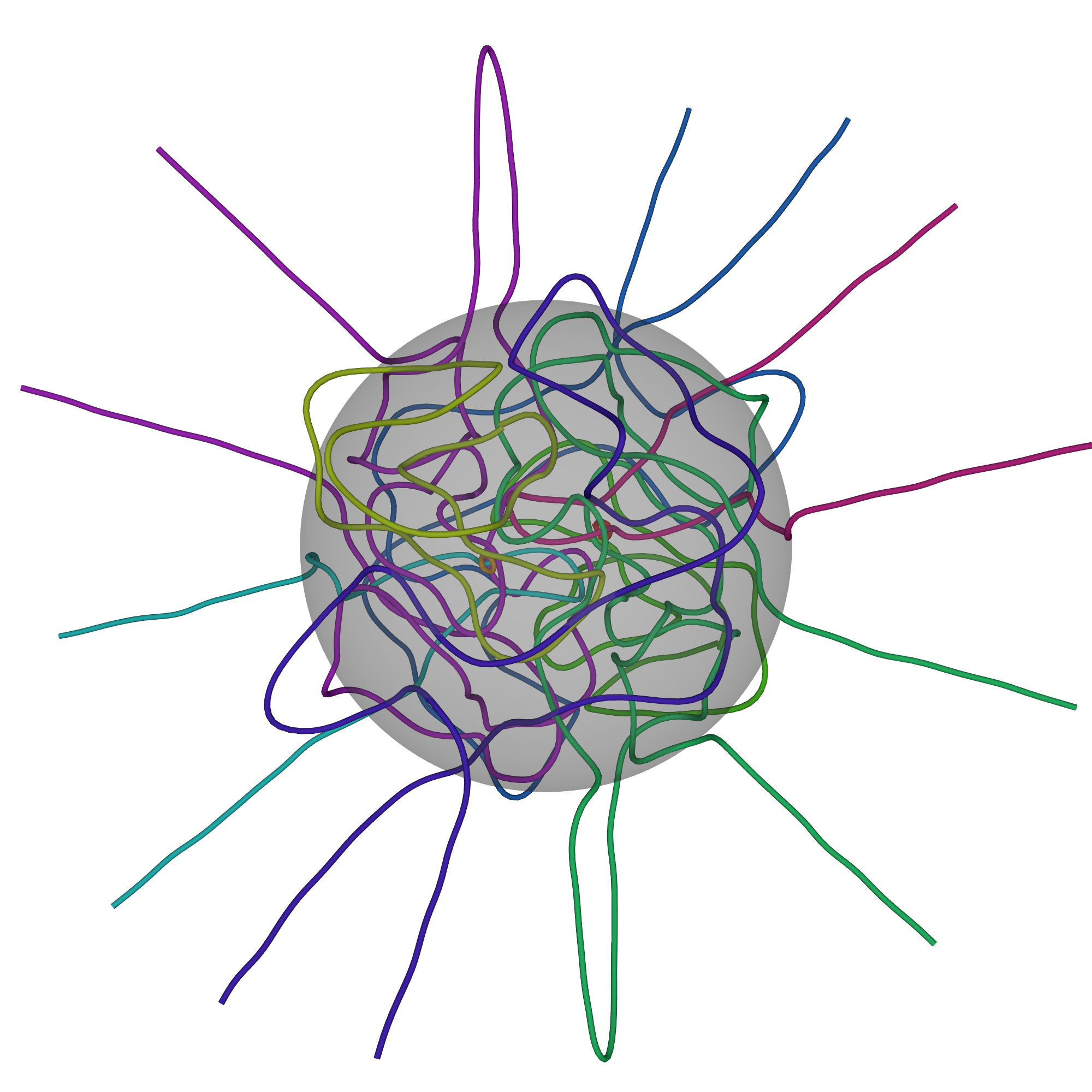}
  }
  \caption{Vortices in a random eigenfunction of the 3DHO with
    N=8. The grey region marks the classical volume where the energy
    is greater than the potential, and which effectively bounds the
    vortex tangle; vortices become straight lines as they move further
    from this shell.}
  \label{fig:qho_classical_region}
\end{figure}

\begin{figure*}
  \centering{
  \includegraphics{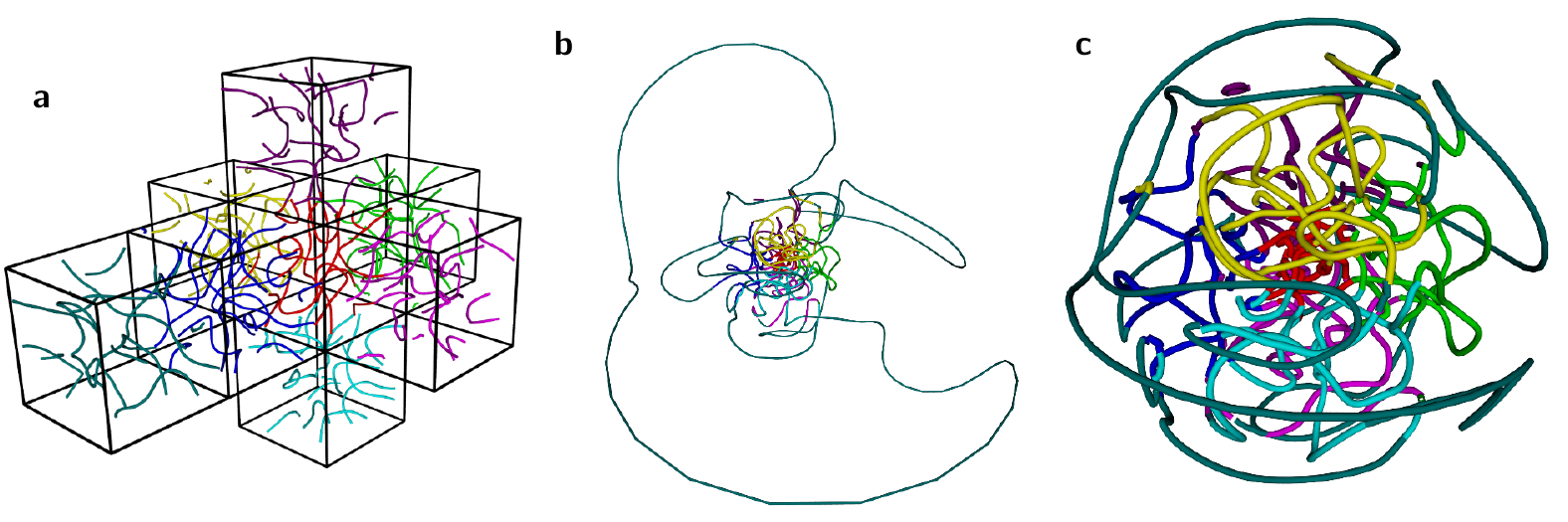}
  }
  \caption{ Projections of the 3-sphere to 3-dimensional Euclidean
    space, illustrated with the vortices of a random eigenfunction
    with $N=7$.  \textbf{a} shows the vortices in each octant of a net
    of the 3-sphere (i.e.~a discontinuous map, with the geometry
    recovered by joining cubic cells along their faces).  The vortices
    within each such octant are given a different colour.  In \textbf{b} the
    same vortices are shown via stereographic projection, continuous
    and angle preserving but significantly distorting distances, with
    the vortices still coloured according to the octants they passed
    through in \textbf{a}.  \textbf{c} shows the rescaled stereographic projection
    used in Figures 1, 3 and 4 of the main text.  }
  \label{fig:sphere_projections}
\end{figure*}

In each system, we compute the total vortex length per random
eigenfunction.  The distribution of these total lengths is found to be
numerically strongly peaked at a value given by the integral over the
volume of the mean vortex density, which is analogous to calculations
of nodal lengths for real random eigenfunctions
\cite{berry02,rudnick08,gnutzmann14,wigman09,bies2002}
(the width of the distribution is in each case proportional to the
square root of the reciprocal of the degeneracy).
Significantly, this total line length is proportional to the systems'
total energy $E_N$.  This justifies our comparison in the main text,
of energy against knotting probability and complexity, as this
provides a physical measure of the total arc length in each tangle.
The reference wavelength $\lambda$ in the main text is proportional to
$E_N^{-1/2}$.  In all cases, we estimate the numerical error in the
calculated total arclength (including that on smoothing the sampled
curves) to be no bigger than $5\%$.  Decreasing this would require
significantly higher resolution in the sampling, and of course would
not affect the topological results.

The vortex density in the ideal isotropic complex RWM is well known to
be $2M E/3\pi\hbar^2$~\cite{F,G}.
The other systems approach this
limit when $N \gg 1$, but in slightly different ways.  The vortex
density in the periodic 3-cell depends weakly on direction~\cite{U},
but at the values of $N$ considered here can be taken to be the
isotropic density.  The density in the 3-sphere is constant, and can
be found using standard methods to be $E_N/3\pi = N(N+2)/3\pi$ (as
above, ignoring physical constants and assuming a 3-sphere of unit
radius), consistent with the isotropic random wave model result.  The
vortex density of the 3DHO is the most complicated, as the density is
inhomogeneous and isotropic (depending on the value of the radius),
and we omit detailed calculations here.  Although the detailed results
depend subtly (although not strongly) on $N$, the total arclength we
calculate (truncating at twice the classical radius) is indeed found
to be proportional to the total energy within comparable error.

In the results of the main text (including Figures 1 and 2), the
calculations involved the 3-torus with $N=9$, with average total
arclength approximately $\num{2000}~\lambda$; the 3-sphere with $N=17$ and average
arclength approximately $\num{1930}~\lambda$; and the 3DHO with $N=21$ and average
arclength approximately $\num{1830}~\lambda$.  Although these are not exact matches, they
are sufficiently close to compare topological statistics which are
representative of general trends.

\section*{Supplementary Note 2:\\ Numerical techniques}
\label{sec:numericaltechniques}

Vortex lines are numerically tracked in 3D complex wavefunctions via a
recursively resampled Cartesian grid method (RRCG).  The core
procedure follows previous numerical
experiments~\cite{vachaspati84,caradoc-davies99,Y}, sampling the field
at points on a 3D Cartesian grid and searching for local 2D grid
plaquettes that are penetrated by a vortex.  Vortices can be located
in this 2D problem either via their intensity (which must be zero) or
the circulation of their phase (which must be $2\pi$ in a path around
the edge of a penetrated plaquette).
It is standard to make use of this latter property, as zeros of the
intensity are difficult to pinpoint numerically whereas the integrated
total change of the phase can be detected even with relatively few
sample points situated far from the vortex core; in fact, it is
possible to detect most vortex penetrations with just the four sample
points at the corners of grid plaquettes with side length around
$0.1\lambda$.

A vortex line is additionally \emph{oriented} by the right-handed
sense of its quantised phase circulation, and according to this
orientation must both enter and leave any grid cell that it passes
through (it cannot simply terminate).  The above procedure therefore
normally detects the passage of a vortex through two different faces
of each 3D grid cell.  The vortex curve is recovered by joining these
points and connecting each line segment with those in neighbouring
cells to build up a piecewise-linear approximation to the
three-dimensional vortex tangle.

This basic procedure does not perfectly detect vortex curves; vortex
penetration of a 2D plaquette may not be detected if the local phase
change is too anisotropic on the scale of the lattice spacing.  This
occurs especially when the lattice spacing is large, or if vortices
approach closely, since in this case the $2\pi$ integrated phase about
multiple vortices cannot be distinguished from the zero phase change
which would mean a vortex is not present.  Such problems give
apparently discontinuous vortex lines, and the numerical procedure
\emph{resamples} the complex wavefield in the cells around these
apparent discontinuities, generating a new local grid with higher
resolution and repeating the search for vortices using this new
lattice.  If a vortex is discontinuous on the new grid, the resampling
procedure is repeated recursively, and is guaranteed to terminate
eventually since at very small lengthscales the smoothness of the
field limits large phase fluctuations.  By matching the different
numerical lattices with one another and joining vortices where they
pass between them, the recovered vortex curves are continuous within
the full numerically sampled region, forming locally-closed loops or
terminating on its boundaries.  A primary advantage of this method is
that it correctly resolves the local topology of vortex lines without
requiring a prohibitively high resolution sampling over the entire
field.  This issue has alternatively been addressed in previous
studies using physical arguments~\cite{Y},
an extra random
choice~\cite{vachaspati84} or a different grid
shape~\cite{hindmarsh95}, but none of these options is so numerically
convenient while guaranteeing robust results.  The resampling
procedure can also be used to enhance the recovery of local vortex
geometry, as described in~\cite{U},
but this is not important to the
topological results described here.

The RRCG algorithm must further be modified in each of the three
different systems of wave chaos we consider.  With periodic boundary
conditions, the finite numerical grid is itself made periodic along
each Cartesian axis of the periodic cell, but the RRCG procedure is
otherwise unaffected.  Vortex loops are recovered by `unwrapping'
vortex segments through the periodic boundaries, equivalent to tiling
space with periodic cells and following each loop continuously until
its starting point, so the net vortex loop can (and often does) pass
through several periodic cell.

In the harmonic oscillator, vortices may extend to infinity and we
only consider vortex length within a finite radius of the origin.  As
distance from the origin increases beyond the classical radius, vortex
curves tend to radial lines without further tangling (clearly visible
in Supplementary Figure~\ref{fig:qho_classical_region}, or Figure 1 of the main text), and we take
the cutoff at twice the classical radius $\sqrt{2E}$.

Tracking vortices in the 3-sphere is more complicated since it
must be projected to flat real space to make it accessible to our
3D Cartesian grid based numerical method. Standard methods such as
stereographic projection are numerically inefficient because they
greatly distort distances and therefore vortex densities, such that an
initial numerical resolution sufficient to detect vortices in the
densest regions will be far higher than necessary in other areas where
the vortex densities are lower.
We instead divide the 3-sphere into a net of eight cubes, with each
taken to be a Cartesian grid covering one of the eight octants of the
3-sphere.  The RRCG algorithm is run on each octant grid, with overall
topology recovered by identification of faces in the overall net. This
process is illustrated in Supplementary Figure
\ref{fig:sphere_projections}, where \ref{fig:sphere_projections}a
shows vortices in each of the cubic octants of the net, discontinuous
where they meet the octant faces, while
\ref{fig:sphere_projections}b-c show the same vortices in two
continuous projections to $\mathbb{R}^3$; these are respectively
stereographic projection and the projection used in the main text.
Although the spatial round metric of the 3-sphere does vary over each
octant, the length variation is in fact relatively small (by no more
than a factor of $2$) and does not significantly impede vortex
tracking efficiency.

\section*{Supplementary Note 3:\\ Topological background and techniques}

\begin{figure}
  \centering{
 \includegraphics[width=0.49\textwidth]{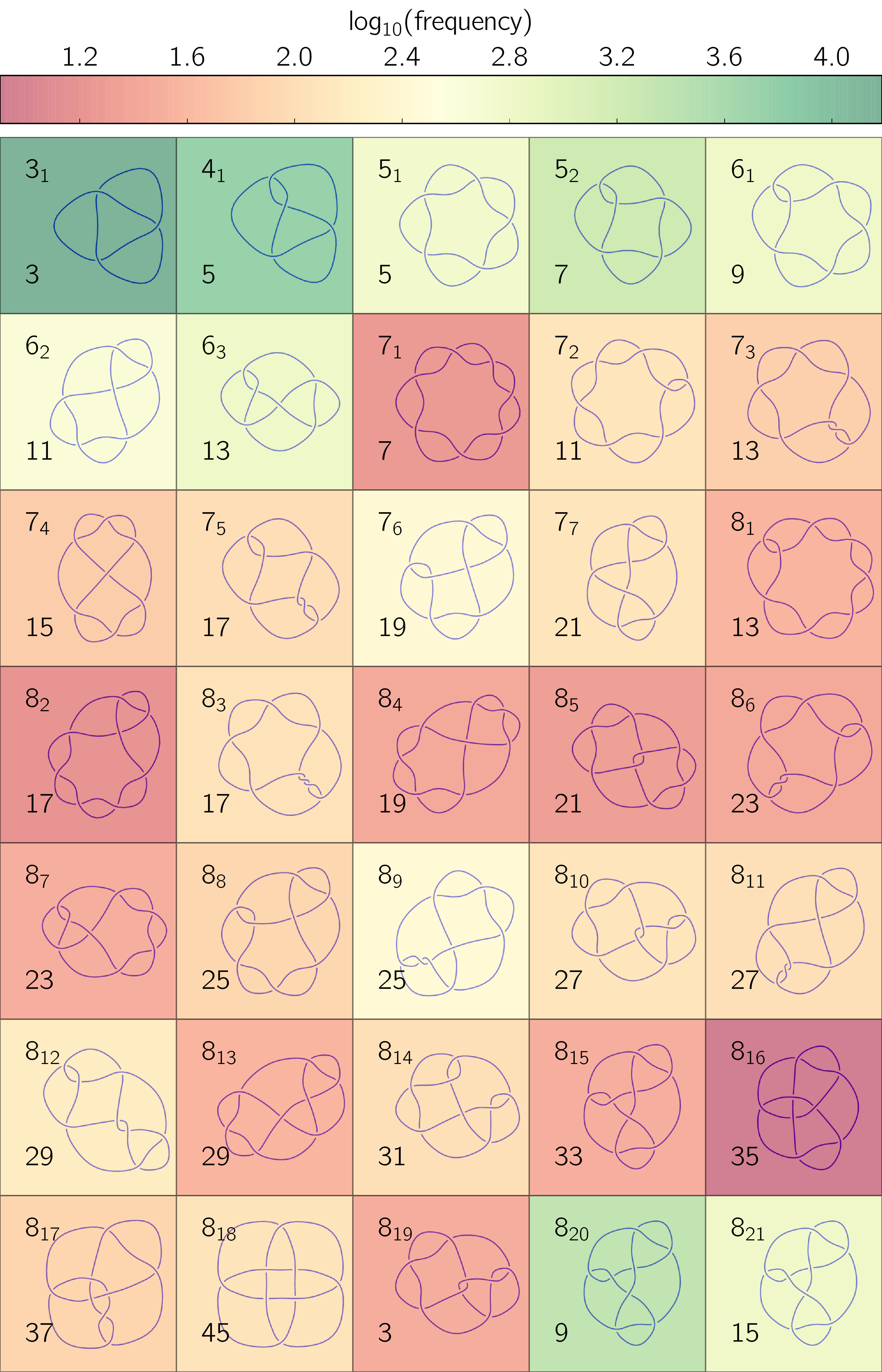}
 }
 \caption{The first 35 non-trivial prime knots, ordered by minimal
   crossing number.  These are all the prime knots with 8 or fewer
   crossings in their minimal projection, with the standard name of
   the knot ($3_1$, $4_1$, etc) in the top left.  The value of the
   knot determinant $|\Delta(-1)|$ for each knot is also given in the
   lower left of each cell; of these, none has the same determinant as
   the unknot (which is unity).  The cells are coloured according to
   the $\log$ of its frequency of occurrence in all of our data
   (unnormalised, and taken over across all systems at various
   different energies), as detected by the Alexander polynomial at
   roots of unity. This figure includes a small number of incorrect
   identifications; the knots $8_{20}$ and $8_{21}$ are
   surprisingly common here, but this is because their Alexander
   polynomials are equal to those of the more common composite knots
   $3_1^2$ and $3_1\#4_1$ (not depicted). These misidentifications
   would be corrected by the hyperbolic volume or Vassiliev
   invariants, which are not applied here. The image of each knot is from the
   KnotInfo Table of Knot Invariants~\cite{knotinfo}, and the
   invariant values from KnotInfo and the Knot Atlas~\cite{X}.
 }
 \label{fig:knot_table}
\end{figure}

The analysis of topology in wave chaos requires that the topological
\emph{knot type} of each vortex curve can be distinguished.  It is
standard to accomplish this through the calculation of \emph{knot
  invariants}, which are mathematical objects (integers, polynomials,
...) that can be computed from the geometric conformation but are the
same for all representations of the same topological knot (i.e.~under
ambient isotopy).  Mathematical knot theory provides many such
invariants, which have been used to develop a taxonomy summarised in
\emph{knot tables}, although no invariant is known to distinguish all
knots.  The values of invariants associated with the \emph{unknot}
(i.e.~a loop which is not knotted) are usually trivial, whereas the
invariants of proper knots usually take other values.  Knot tables are
usually ordered by the invariant \emph{minimal crossing number},
i.e.~according to the smallest number of \emph{crossings} the knot
admits on projection into a 2-dimensional plane; the first few prime
knots are written $3_1$ (the only knot with minimal crossing number 3,
i.e.~the trefoil knot), $4_1$ (the only knot with minimal crossing
number 4, i.e.~the figure-8 knot), then $5_1, 5_2$, and so on. Supplementary Figure
\ref{fig:knot_table} shows the 35 non-trivial prime knots with 8 or
fewer minimal crossings.  \emph{Composite knots}, those which can be
separated into distinct prime knot components each with smaller
minimal crossing number such as Figure 1e in the main text, are not
included in this table. Composite knots are referred to as
combinations of prime knots joined by $\#$, or by exponents for a
repeated component; for instance, $3_1^2$ for the double trefoil knot or
$3_1\#4_1$ for the join of the first two non-trivial knots.
Invariants of composite knots can usually be factorised in some sense
into those of the component knots, and they are denoted by adjoining
the notation of their prime components.  Whether a given vortex curve
is knotted is found by calculating one or more knot invariants, and
then looking these up in the knot table.  The values of invariants can
also give information about different families and classes of which a
given knot is a member~\cite{X,knotinfo}.

Our primary requirement is to distinguish lines that are knotted from
those that are not, and to be able to sort knotted curves by some
simple knot invariant which measures their complexity.
For this we employ the \emph{Alexander polynomial}~\cite{G}
$\Delta(t)$, which is straightforward to calculate up to an unknown
factor of $t^n$~\cite{k}
(e.g.~the unknot has Alexander polynomial
$\Delta(t) = t^n$, the trefoil knot $3_1$ has
$\Delta(t) = (1-t+t^2)t^n$) as the determinant of a matrix with
dimension $n-1$ for projection of the knot with $n$ crossings.  Some
knotted curves have Alexander polynomial $\Delta(t) = t^n$, like the
unknot, but such knots are comparatively rare; the simplest examples
are two knots with minimal crossing number 11, already complex enough
to be highly uncommon in eigenfunction vortex tangle.  Supplementary Figure
\ref{fig:knot_table} demonstrates this trend; each knot is coloured by
the frequency of its occurrence across all eigenfunction data in all
different systems, with most knots by frequency having low minimal
crossing numbers, and those with $8$ crossings already being up to
\num{1000} times less common.

Since the projections of curves in our numerical tangle may have
several thousand crossings (even after algorithmic simplification), it
can be impractical to calculate the full Alexander polynomial
symbolically.  Thus we evaluate the Alexander polynomial at specific
values, conveniently the first three nontrivial roots of unity, $-1$
(giving the \emph{knot determinant}~\cite{k}),
 $\exp(2\pi i/3)$ and
$i$.
The absolute value $|\Delta(t)|$ is invariant under this substitution
regardless of the unknown factor of $t$, and this combination of
integer values discriminates the tabulated knots almost as well as the
symbolic Alexander polynomial itself; for instance, they are equally
discriminatory when distinguishing the 802 prime knots with 11 or
fewer crossings.  The determinant is a commonly used tool for
identifying knots in numerical studies~\cite{moore04,k}, although we
have not found $\Delta$ at other roots of unity used elsewhere in
numerical knot identification.

The knot determinant is also convenient on its own as a measure of
knot complexity; it takes its minimum value on the unknot,
$\Delta(-1)=1$, and tends to increase with crossing number (we find it
appears on average linearly related to the exponential of the minimal
crossing number, consistent with known bounds~\cite{stoimenow2003}),
and for a composite knot is the product of determinants of its prime
components~\cite{soteros92}.  Many other invariants fulfil these
conditions, but the determinant is convenient due to its ease of
calculation, the same reason that it is used already in knot
detection.
Supplementary Figure \ref{fig:knot_table} includes this complexity trend for each of
the non-trivial knots with 8 or fewer minimal crossings, taking values
from $3$ (for the simple trefoil knot $3_1$) to $45$ (for $8_{18}$).

Where it is necessary to distinguish the knot type beyond the
discriminatory ability of the Alexander polynomial, such as in Figure
1 of the main text, further invariants
are used.  Modern knot theory supplies many powerful options, but most
of these are impractical to calculate rapidly for large numbers of
geometrically complex projections (being calculable only in
exponential time), and we instead use more efficient options that are
nevertheless sufficient.  First, the \emph{Vassiliev invariants} of
order two and three are also integer invariants of knots, practically
calculable in square or cubic time respectively in the number of
crossings of a given diagram, but adding further discriminatory power
beyond that of the Alexander polynomial~\cite{W}.
 These invariants
have been used previously in numerical knot
identification~\cite{moore04}.  We also use the \emph{hyperbolic
  volume}, which takes values in the real numbers and is nonzero only
for the so-called \emph{hyperbolic knots}, but is highly
discriminatory among this class, and for this reason has seen major
use in knot tabulation~\cite{hoste1998}. Most
tabulated prime knots are hyperbolic (of
the the 1.7 million prime knots with 16 or fewer crossings, only 32
are not~\cite{hoste1998}), but composite knots always have volume zero
and so are also readily separated from prime knots in this way.  The
hyperbolic volume is calculated using the standard topological
manifold routines in SnapPy~\cite{V},
which return only an
approximation but are reliable over the range of complexities we
address.

Neither these Vassiliev invariants nor the hyperbolic volume are
perfect knot invariants, but combined with the Alexander polynomial
they are sufficiently discriminatory to unambiguously identify most
simple knots where necessary, such as those in Figures 1 and 4 of the
main text.  They also further verify that in practice the Alexander
polynomial rarely fails to detect knotting among the vortex lines of
our eigenfunction systems, as those prime knots with $\Delta$
indistinguishable from the unknot are generally easily detected to be
hyperbolic.

\section*{Supplementary Note 4:\\ Symmetries of knots}

Some aspects of knotting in eigenfunctions are dominated by the
\emph{symmetry} of the system.  Such symmetries have already been
removed in our analysis of wave chaos under periodic boundary
conditions, but remain in the eigenfunctions of both the 3-sphere and
3DHO.

In the 3-sphere, all eigenfunctions satisfy the condition
$\Psi(\psi, \theta, \phi) = (-1)^N\Psi(\pi - \psi, \pi - \theta, \phi +
\pi)$
and, since, vortices are nodal lines, a vortex at a given position is
always paired with a vortex at its 3-sphere antipode. If these points are on
different vortex lines then both lines are identical up to a rotation
of the 3-sphere by $\pi$ through some plane in four dimensions, and so
they have the same knot type.  If two antipodal points are positions
on the same vortex line then the entire vortex curve must be symmetric
(carried to itself) under this rotation. Not all knot types can meet
this geometric constraint; those that can do so are a subclass of
knots that are \emph{strongly amphicheiral}.

Strong amphicheirality is an extension of the more commonly considered
chirality of knots; a knot is chiral if it is not equivalent to its
mirror image (i.e.~with the overstrand and understrand at each
crossing switched), and otherwise is called amphicheiral (or
equivalently achiral). Both types are common, e.g.~the trefoil knot is
chiral (shown in Supplementary Figure \ref{fig:symmetric_knots}a), but the next
non-trivial knot, $4_1$, is amphicheiral~\cite{adams94} (evident in
the projection of Supplementary Figure \ref{fig:symmetric_knots}c, equal to its
mirror image under a rotatation by $\pi$ about the marked point,
discussed further below). Strong amphicheirality additionally demands
that the knot be equivalent to its mirror not just in its knot type,
but under a specific involution of the 3-sphere, i.e.~a geometric
transformation that is its own inverse~\cite{kawauchi96}. Supplementary Figure
\ref{fig:symmetric_knots}b-d shows three example diagrams of
strongly amphicheiral knots in which the involution is rotation in two
dimensions by $\pi$ about the marked point, which in each case takes
the knot diagram to its mirror image; b gives an example of how any
composite of a knot with its mirror image admits a strongly
amphicheiral conformation~\cite{kawauchi96}, c shows strong
amphicheirality
of the knot $4_1$, and d the same for the more complicated
$10_{99}$.

Strongly amphicheiral knots are additionally split into two classes
depending on how the involution affects the orientation of the curve,
which for an arbitrary curve may not matter but in vortex lines is
fixed by the orientation of the phase. A knot is \emph{strongly negatively
  amphicheiral} if this orientation is preserved under the involution,
or \emph{strongly positively amphicheiral} if its orientation is
reversed. Supplementary Figure \ref{fig:symmetric_knots}b and \ref{fig:symmetric_knots}c show strongly
negatively amphicheiral examples (note the reversal of the marked
orientation under rotation about the marked point), and this is
additionally the reason for these diagrams being drawn as open curves
closing at infinity; under the involution of rotation, the origin of
rotation is privileged such that the curve must pass through this
point and close at infinity (equivalent to the antipode considered on
the 3-sphere). No other conformation would be able to meet the strong
negative amphicheiral symmetry.
In contrast, Supplementary Figure \ref{fig:symmetric_knots}d shows a strong
positive amphicheiral conformation of the knot $10_{99}$. In fact,
$10_{99}$ is the simplest knot with strong positive amphicheiral
symmetry, and (with a different conformation) also supports strong
negative amphicheirality.

In eigenfunctions of the 3-sphere, the symmetry under rotation
reverses the local vortex line orientation according to its phase
circulation. This means that such vortices passing through antipodal
points can only form strong positive amphicheiral knots, of which the
simplest example is the unknot but the first non-trivial prime example
is $10_{99}$. Although not discussed in the main text, this knot and
others with the same symmetry occur with disproportionate frequency in
3-sphere eigenfunctions, with simpler knots occurring only as
symmetric antipodal pairs (under the eigenfunction symmetry) or as
strong positive amphicheiral composite knots. This symmetry also has
an equivalent effect at all $N$, and so does not lead to significant
patterns in knotting probability with energy, as can be seen in Figure
3a of the main text.

\begin{figure}
 \centering{
 \includegraphics{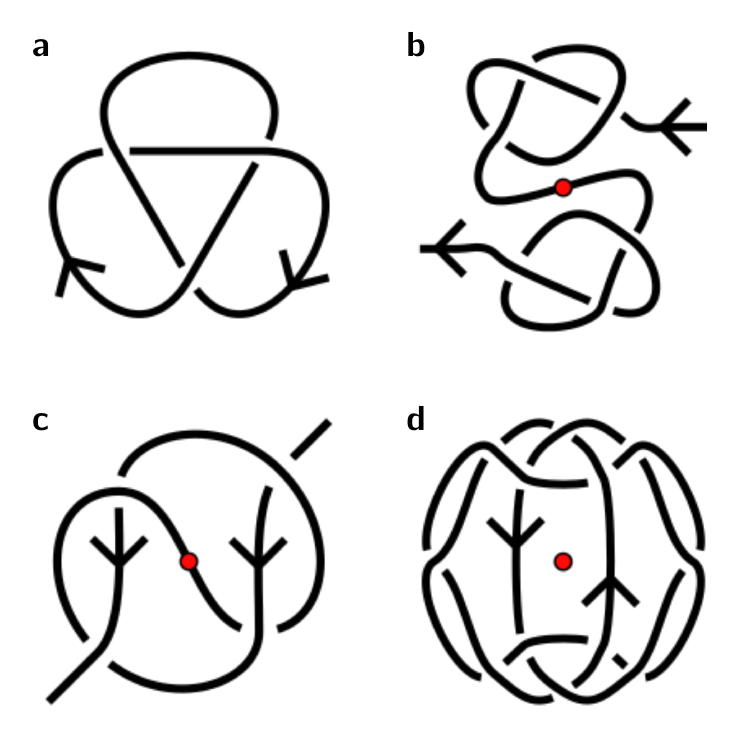}
 }
 \caption{Knots exhibiting different symmetries.  In \textbf{a} a trefoil
   knot, which is chiral and cannot be transformed to its mirror image
   without passing the curve through itself.  \textbf{b}-\textbf{d} show strongly
   amphicheiral conformations of three other knots, equivalent to
   their mirror images under a rotation by $\pi$ about the marked red
   point, and with or without an orientation reversal of the line; in
   \textbf{b} a strongly negatively amphicheiral composite double trefoil knot
   ($3_1^2$), in \textbf{c} a strongly negatively amphicheiral figure-eight
   knot ($4_1$), and in \textbf{d} a strongly positively amphicheiral knot
   $10_{99}$.}
 \label{fig:symmetric_knots}
\end{figure}

Eigenfunctions of the 3DHO have a similar symmetry under inversion
through the origin, $\Psi(\vec{r}) = (-1)^N\Psi(-\vec{r})$, but with
the difference now that this supports only strong negative
amphicheiral symmetry; the vortex tangent direction is preserved under
the inversion.  As with the 3-sphere, any vortex line passing through
$\vec{r}$ is paired with one at $-\vec{r}$, and if these points are on
different vortices then the entire vortex line appears twice. If the
points are on the same vortex line then it must take up a strongly
negatively amphicheiral conformation.  Under inversion, the only way to
do so while meeting the symmetry of the eigenfunction is to pass
through the origin and to close at infinity, taking a conformation
such as those in Supplementary Figure \ref{fig:symmetric_knots}b and \ref{fig:symmetric_knots}c. Unlike
in the 3-sphere, only one vortex line in a given eigenfunction can do
so, and at most one strongly amphicheiral knot can appear. Since the
3DHO naturally supports vortex lines which eventually extend to
infinity in straight lines outside the classical radius, it is
possible for the privileged origin vortex line to do so and to be
knotted.

The probability of a vortex passing through the origin depends
directly on $N$; when $N$ is even, random degenerate eigenfunctions
are non-zero at the origin, whereas when $N$ is odd the origin is
always a nodal point and sits on a vortex line. This is the reason for
the observed odd-even discrepancy in knotting probability with energy
in Figure 3 of the main text; when $N$ is even there is no strongly
negatively amphicheral vortex line and all knots occur in pairs of
antipodal mirrors. When $N$ is odd, such a vortex line always exists
and can form a strongly negatively amphicheiral conformation of a
knot. This is relatively common, as the total arclength required to
form such a knot is often lower than for the two symmetric copies
otherwise required. This knot is frequently prime (unlike with strong
positive amphicheirality, there are several strongly negatively
amphicheiral knots with fewer than 10 crossings), but can also be a
composite of a knot with its mirror image. Compatible knots are thus
overrepresented in the statistics of knot type within the system, but
now only when $N$ is odd explaining the strong parity dependence of
knotting in the 3DHO. The effect is also strong enough to persist even
at $N$ high enough that there is a 50\% or higher chance for one or
more pairs of the other vortex lines in a given eigenfunction to form
knots.

\end{document}